\documentclass[aip,pop,graphicx]{revtex4-1}
\usepackage{graphicx}
\usepackage{amsmath}

\begin{document}
\title{Computing the Paschen curve for argon with speed-limited particle-in-cell simulation}
\author{Joseph G. Theis}
\email[]{joseph.theis@colorado.edu}
\affiliation{Center for Integrated Plasma Studies, University of Colorado, Boulder, Colorado 80309, USA}
\author{Gregory R. Werner}
\affiliation{Center for Integrated Plasma Studies, University of Colorado, Boulder, Colorado 80309, USA}
\author{Thomas G. Jenkins}
\affiliation{Tech-X Corporation, 5621 Arapahoe Avenue Suite A, Boulder, Colorado 80303, USA}
\author{John R. Cary}
\affiliation{Center for Integrated Plasma Studies, University of Colorado, Boulder, Colorado 80309, USA}
\affiliation{Tech-X Corporation, 5621 Arapahoe Avenue Suite A, Boulder, Colorado 80303, USA}
\date{\today}
\begin{abstract}
Upon inclusion of collisions, the speed-limited particle-in-cell (SLPIC) simulation method successfully computed the Paschen curve for argon. The 1D3V simulations modelled an electron cascade across an argon-filled capacitor, including electron-neutral ionization, electron-neutral elastic collisions, electron-neutral excitation, and ion-induced secondary electron emission. In electrical breakdown, the timescale difference between ion and electron motion  makes traditional particle-in-cell (PIC) methods computationally slow. To decrease this timescale difference and speed up computation, we used SLPIC, a time-domain algorithm that limits the speed of the fastest electrons in the simulation. The SLPIC algorithm facilitates a straightforward, fully-kinetic treatment of dynamics and collisions. SLPIC was as accurate as PIC, but ran up to 200 times faster. SLPIC accurately computed the Paschen curve for argon over three orders of magnitude in pressure.
\end{abstract}
\maketitle

\section{Introduction} \label{sec:intro}
Townsend discharge is the conduction of electric current in a gas due to a cascade of ionizing collisions.\cite{townsend1900conductivity} In typical discharge experiments, the gas is enclosed between the electrodes of a capacitor, and there is an initial source of free electrons. The electrons accelerate in an applied electric field and ionize the neutral atoms, creating more electrons, which ionize more atoms. The ions accelerate towards and impact the cathode, where they can produce secondary electrons. When the combined effects of ionizing collisions and ion-induced cathode emission yield more electrons at the cathode than initially seeded, a self-sustaining discharge is triggered. The voltage difference at which this occurs is the breakdown voltage. 

The breakdown voltage $V_b$ of a gas-filled capacitor is (to a good approximation) a function of $pd$, where $p$ is the pressure of the gas and $d$ is the distance between the electrodes. A Paschen curve is the relationship between $V_b$ and $pd$ for a given gas.\cite{paschen1889ueber} Paschen curves have a minimum $V_b$ at some intermediate $pd$ and higher $V_b$ at the extremes (this shape is shown later in Fig.~\ref{fig:paschen}). This characteristic shape is due to the competition between ionizing collisions, which multiply the number of free electrons, and non-ionizing collisions, which prevent electrons from gaining enough energy to ionize atoms. At low $pd$, the ionization mean free path is too long for a sufficient number of ionizations to occur during crossing. At high $pd$, the non-ionizing mean free path is too short for electrons to reach the ionization energy. Paschen curves will be developed further in \S\ref{sec:paschen}.

Electrical discharge and breakdown have been extensively studied both for a multitude of applications, including lighting,\cite{sugiura1993review} plasma processing,\cite{lieberman2005principles} and lasers,\cite{belasri1993cathode} as well as to avoid detrimental effects, including damage to electrical devices,\cite{gao2003improved} explosions in chemical processing facilities,\cite{gillman2007mechanical} and damage to aerospace systems.\cite{ashwell1966high} Following Townsend's description of the mechanism behind gas discharge,\cite{townsend1900conductivity} various models and simulation methods have been employed to study the phenomena.\cite{phelps1999cold} Analytical models have proven effective for computing breakdown and steady-state current amplification in simple geometries;\cite{phelps1999cold} however, understanding the particle dynamics of discharge and handling complex geometries requires simulation. 

Particle-in-cell (PIC) and fluid methods have been the primary simulation techniques applied to model gas discharges.\cite{boeuf1991pseudospark,lieberman2005principles} Fluid approaches evolve the mean properties of a particle species and can therefore handle situations involving many particles and mean free paths that are small compared with the system size. However, the fluid approaches assume Maxwellian particle velocity distributions, while discharges often have non-Maxwellian distributions.\cite{boeuf1991pseudospark} 

PIC methods evolve the distribution function $f(x, v, t)$ in time via the method of characteristics and, therefore, can simulate arbitrary distribution functions.\cite{birdsall1991particle} Collisions can be included in PIC using the PIC Monte-Carlo Collisions framework (PIC-MCC).\cite{birdsall1991particle,bird1994molecular,nanbu2000probability} PIC-MCC will be discussed further in \S\ref{sec:collisions}. PIC-MCC requires that the smallest mean free path be resolved by the grid cell size $\Delta x$. Further, in PIC, the timestep $\Delta t$ is limited according to $\Delta t \lesssim \Delta x/v_{\rm max}$, where $v_{max}$ is the maximum particle speed. In gas discharge simulations, the large mass difference between the ions and the electrons results in ion timescales much greater than the timestep imposed by the maximum electron speed. This makes PIC-MCC simulations computationally slow. Because PIC-MCC is slow, variations of PIC simulation have been applied to discharge. One approach is treating the ions as immobile to eliminate the timescale mismatch,\cite{bagheri2018comparison,sun2018simulation} but this technique fails when ion motion becomes relevant as is the case for Townsend discharge. Another technique is electron sub-cycling, in which the timestep for the electron is smaller than that of the ion,\cite{adam1982electron} but this technique yields at most a factor of two speed-up. Hybrid methods combine PIC and fluid approaches (e.g., with fluid electrons but PIC ions),\cite{boeuf1991pseudospark} but they fail to capture the full velocity distribution as PIC does.
 
In this work, we apply the speed-limited particle-in-cell (SLPIC) simulation method\cite{werner2018speeding} to electrical breakdown in argon. SLPIC is a time-domain algorithm that limits the speed of the fastest particles in the simulation to enable larger time steps and, therefore, faster computing times. The SLPIC algorithm facilitates a straightforward, fully-kinetic treatment of effects such as secondary emission at the cathode and collisions. A detailed introduction to SLPIC is provided in \S\ref{sec:slpicIntro}. This work is the first demonstration of the integration of SLPIC particles with PIC-MCC, showing that speed-limiting techniques can be used in collisional low-temperature plasma discharges. We chose to simulate electrical breakdown, and, specifically, Paschen's law, because multiple collision mechanisms must be correctly simulated to compute the breakdown voltage accurately. We were also able to quantify the speed-up in runtime of the SLPIC simulations relative to the regular PIC simulations.

In this paper, we begin with a review of Paschen curves and SLPIC. We then introduce the SLPIC Monte Carlo Collisions framework in \S\ref{sec:collisions}. The simulation design and parameters will be provided along with our methodology for determining breakdown in \S\ref{sec:method}. In \S\ref{sec:results}, we compare our simulation results (a Paschen curve) with experimental data and quantify the computational speed-up of the SLPIC algorithm relative to normal PIC.

\section{Background} 
\subsection{Paschen Curve} \label{sec:paschen}
Electrical breakdown in a gas-filled capacitor is caused by the ionization of the neutral gas by electrons and the secondary emission of electrons at the cathode due to ion impact. The average number of ions produced per electron per unit length travelled is termed\cite{townsend1900conductivity} Townsend's first ionization coefficient $\alpha$. The average number of secondary electrons emitted per ion impact is Townsend's second ionization coefficient $\gamma_{se}$. The coefficient $\alpha$ depends on the voltage $V$ across the gap and the collision cross-sections. The coefficient $\gamma_{se}$ depends on the ion species and the electrode material and treatment. When $\gamma_{se}(e^{\alpha d}-1) > 1$ breakdown occurs. 

Paschen curves relate the breakdown voltage $V_b$ of a gas-filled capacitor to the product of the gas pressure $p$ and the distance $d$ between the electrodes.\cite{paschen1889ueber} Paschen curves are not monotonic due to a competition between electrons gaining energy in the electric field and losing energy in collisions. Paschen curves have a characteristic minimum at an intermediate $pd$. At low $pd$, most electrons stream through the capacitor without ionizing enough atoms. At high $pd$, few electrons reach the threshold ionization energy due to frequent non-ionizing collisions. Paschen's law is a function relating $V_b$ to $pd$:
\begin{equation} \label{eq:paschen}
V_b = \frac{Bpd}{\ln(Apd)-\ln[\ln(1+1/\gamma_{se})]}.
\end{equation}
Paschen's law assumes that, except for the initial seed electrons, ion impact at the cathode and electron-neutral ionizations are the only sources of electrons, and that $\alpha$ can be expressed as:
\begin{equation} \label{eq:alpha}
\frac{\alpha}{n_g}=kTA\exp\left(\frac{-kTBn_g}{E}\right)
\end{equation}
where $kT$ is the thermal energy, $n_g$ is the neutral gas density, and $E=V/d$ is the constant electric field. $A$ and $B$ are related to the collision cross-sections and are determined by fitting Eq.~(\ref{eq:alpha}) (which gives the field-intensified ionization cross-section $\alpha/n_g$ as a function of the reduced field $E/n_g$) to the experimental results for a given gas. In Paschen's law, $A$ and $B$ are assumed to be constant for a given gas species. Paschen's law will be plotted in Fig.~\ref{fig:paschen}, with the values of $A$, $B$, and $\gamma_{se}$ taken from \citet{lieberman2005principles}.

Different gases yield different Paschen curves due to different cross-sections, ionization energies, and secondary emission yields at the cathode. In this work, we simulated a neutral argon gas. Argon has been studied extensively in the context of electrical breakdown, including simulation,\cite{radmilovic2005modelling, kutasi2000hybrid, boeuf1991pseudospark} and experimental work.\cite{mariotti2004experimental, yamabe1983measurement, phelps1999cold} Research has been focused on identifying the mechanisms responsible for breakdown and the regimes in which these mechanisms are relevant. In this work, we will compare our simulation results to experimental results compiled by \citet{phelps1999cold}.

\subsection{SLPIC} \label{sec:slpicIntro}
Before describing how collisions can be implemented within the SLPIC method in \S\ref{sec:collisions}, we review the derivation of SLPIC, following \citet{werner2018speeding}. PIC numerically evolves a particle distribution $f(\mathbf{x}, \mathbf{v}, t)$ according to the Vlasov equation: 
\begin{eqnarray} \label{eq:Vlasov}
  \partial_t f(\mathbf{x}, \mathbf{v}, t) 
  &=&
  -\mathbf{v} \cdot \partial_{\bf x} f(\mathbf{x}, \mathbf{v}, t) 
  -\mathbf{a} \cdot \partial_{\bf v} f(\mathbf{x}, \mathbf{v}, t) 
,\end{eqnarray}
where the acceleration $\mathbf{a}=\mathbf{a}(\mathbf{x}, \mathbf{v}, t)$ is typically determined by electromagnetic fields. SLPIC numerically evolves $f(\mathbf{x},\mathbf{v},t)$ according to an \emph{approximate} Vlasov equation. We multiply Eq.~(\ref{eq:Vlasov}) by a yet-to-be-chosen function $\beta(v)$ (where $v=\|\mathbf{v}\|$), which will turn out to limit particle speeds,
\begin{eqnarray} \label{eq:VlasovBeta}
  \beta \partial_t f(\mathbf{x}, \mathbf{v}, t) 
  &=&
  -\beta \mathbf{v} \cdot \partial_{\bf x} f(\mathbf{x}, \mathbf{v}, t) 
  -\beta \mathbf{a} \cdot \partial_{\bf v} f(\mathbf{x}, \mathbf{v}, t) 
,\end{eqnarray}
and then make the critical ``speed-limiting'' approximation:
\begin{eqnarray} \label{eq:slpicApprox}
  \beta \partial_t f &\approx & \partial_t f
.\end{eqnarray}
This approximation is valid if either (1) $\partial_t f(\mathbf{x},\mathbf{v},t) \approx 0$ or (2) $\beta(v) \approx 1$.\footnote{This approximation is subtly different from that in \citet{werner2018speeding}, but only when $\beta$ has an explicit time-dependence---a regime that has not yet been investigated and is irrelevant for this paper.} Thus we arrive at the ``speed-limited'' approximation to the Vlasov equation
\begin{eqnarray} \label{eq:slpicEq}
  \partial_t f(\mathbf{x}, \mathbf{v}, t) 
  &=&
  -\beta \mathbf{v} \cdot \partial_{\bf x} f(\mathbf{x}, \mathbf{v}, t) 
  -\beta \mathbf{a} \cdot \partial_{\bf v} f(\mathbf{x}, \mathbf{v}, t) 
.\end{eqnarray}
In the special case of a steady state (i.e., $\partial_t f\equiv 0$), the speed-limiting approximation is exact and SLPIC, though different from PIC, is as accurate as PIC. For the special choice of $\beta(v)\equiv 1$, SLPIC is exactly the same as PIC. 

Just as PIC simulation evolves the Vlasov equation via the method of characteristics, SLPIC evolves Eq.~(\ref{eq:slpicEq}). We posit a solution that is a sum over macroparticles $p$ that follow trajectories $[\mathbf{x}_p(t), \mathbf{v}_p(t)]$ with weights $w_p(t)$ (a macro-electron with weight $w_p$ represents $w_p$ electrons; i.e., it has mass $w_p m_e$ and charge $-w_p e$):
\begin{eqnarray} \label{eq:Ansatz}
  f(\mathbf{x}, \mathbf{v}, t) 
  &=&
  \sum_p w_p(t) S[\mathbf{x}-\mathbf{x}_p(t)] 
             \delta^3 [\mathbf{v}-\mathbf{v}_p(t)]
\end{eqnarray}
where $S$ is the particle shape function\cite{birdsall1991particle} and $\delta $ the Dirac delta function. Unlike PIC, SLPIC requires the particle weight to vary in time. Plugging Eq.~(\ref{eq:Ansatz}) into Eq.~(\ref{eq:slpicEq}), we find that it yields a solution if:
\begin{eqnarray} 
  \label{eq:slpicEOMx}
  \dot{\mathbf{x}}_p &=& \beta(v_p) \mathbf{v}_p
  \\
  \label{eq:slpicEOMv}
  \dot{\mathbf{v}}_p &=& 
    \beta(v_p) 
    \mathbf{a}(\mathbf{x}_p, \mathbf{v}_p, t)
    \\
   \dot{w}_p &=& w_p [\partial_{\bf x} \cdot (\beta \mathbf{v})
                   + \partial_{\bf v} \cdot (\beta \mathbf{a})]
     = \beta w_p [\partial_{\bf x} \cdot \mathbf{v}
                   + \partial_{\bf v} \cdot \mathbf{a}]
       + w_p \mathbf{a} \cdot \partial_{\bf v} \beta
     %= \frac{w_p}{\beta} \dot{\beta}
     = w_p  \dot{\beta} / \beta 
     \label{eq:slpicEOMw}
\end{eqnarray}
where we have assumed that $\partial_{\bf x} \cdot \mathbf{v} + \partial_{\bf v} \cdot \mathbf{a} = 0$---i.e., the true motion must be phase-space-volume preserving (Liouville's theorem proves this for any Hamiltonian motion). We have also substituted 
$\dot{\beta} \equiv (d/dt) \beta(v_p(t))= \beta\mathbf{a}\cdot \partial_{\bf v} \beta$.

When $\beta(v_p)=1$, these are the usual (PIC) equations of motion. When $0 < \beta(v_p) < 1$, a SLPIC macroparticle, representing physical particles with true velocity $\mathbf{v}_p$, travels with pseudo-velocity $\dot{\bf x}_p=\beta(v_p) \mathbf{v}_p$ and pseudo-acceleration $\dot{\bf v}_p=\beta(v_p)\mathbf{a}(\mathbf{x}_p(t),\mathbf{v}_p(t),t)$. The SLPIC macroparticle follows the same path through phase space as the physical particles it represents, but slower; e.g., in a short time~$\Delta t$, the macroparticle moves $({\bf v}_p,{\bf a})\beta \Delta t$, while a physical particle moves $({\bf v}_p,{\bf a}) \Delta t$. For example, a SLPIC electron in a uniform magnetic field will have the correct gyroradius, but its gyrofrequency will be (unphysically) reduced by a factor~$\beta_{v_p}$.

Choosing $\beta(v_p)<1$ invokes the SLPIC approximation [Eq.~(\ref{eq:slpicApprox})], reducing accuracy; however, it limits the simulated speed of macroparticles from $v_p$ to $\dot{x}=\beta(v_p) v_p$, which allows larger timesteps, hence faster simulation. In most cases, the PIC timestep must be smaller than $\Delta t \lesssim \Delta x/v_{\rm max}$, where $\Delta x$ is the grid cell size and $v_{\rm max}$ is the maximum particle speed. With SLPIC, a larger timestep is allowed: $\Delta t \lesssim \Delta x/[\beta(v_{\rm max}) v_{\rm max}]$. Typically, we choose a speed limit $v_0 \ll v_{\rm max}$, and then define $\beta(v_p) \equiv v_0/v_p$ to ensure that for all particles, $\|\dot{\bf x}\| \leq v_0$, so that we can use $\Delta t \approx \Delta x /v_0\gg \Delta x/v_{\rm max}$. For particles with $v_p<v_0$, however, we define $\beta(v_p)\equiv 1$ so that they experience ``true'' motion. In this way, we speed-limit only the particles (with $v_p>v_0$) that need to be speed-limited to allow a large timestep, $\Delta t=v_0/\Delta x$. It turns out that speed-limiting also reduces the plasma frequency so that it is resolved by the large $\Delta t$ and poses no instability.\cite{werner2018speeding,jenkins2021Dispersion} For simplicity, we will use this specific $\beta(v_p)$ for the rest of the paper: $\beta(v_p)=v_0/v_p$ for $v_p\geq v_0$, and $\beta(v_p)=1$ for $v_p<v_0$, or, equivalently:
\begin{eqnarray} \label{eq:betaAbrupt}
  \beta(v_p) & \equiv & \Theta(v_p-v_0)v_0/v_p + \Theta(v_0-v_p)
\end{eqnarray}
where $\Theta$ is the Heaviside step function (and $v_p=\|\mathbf{v}\|$).

As a macroparticle accelerates so that $v_p>v_0$, its simulated speed $\dot{x}_p=\beta v_p$ remains at $v_0$, and its weight $w_p(t)$ decreases. The decrease in $w_p(t)$ reflects the fact that the physical particle density represented by the macroparticle at time~$t$ decreases, and the macroparticle represents fewer physical particles at time~$t+\Delta t$. In this way, SLPIC can correctly simulate a steady-state beam within which particles accelerate such that their density decreases (like cars exiting a steady-state traffic jam). In SLPIC, a beam macroparticle with $v_p>v_0$ experiences an acceleration and its true velocity $v_p$ increases, but its pseudo-velocity $\dot{x}_p=\beta v_p=v_0$ does not, so the density of macroparticles throughout the beam remains uniform. The decrease in physical density thus results from the decrease in macroparticle weight, and not (as in PIC) from the decrease in macroparticle density. While SLPIC may sometimes result in a more uniform macroparticle density, this is an unintended bonus and it may not always be the case. SLPIC is orthogonal to techniques for managing macroparticles weights/numbers for efficiency and accuracy.\cite{teunissen2014controlling,lapenta1995control}

Therefore, as $v_p$ increases above $v_0$, $\beta(v_p)$ decreases, and so too must $w_p$ decrease. Equation~(\ref{eq:slpicEOMw}) shows that $w_p/\beta(v_p)$ is constant over any particle trajectory; we define this constant to be $w_{j,p}\equiv w_p/\beta(v_p)$. Here we choose the subscript~$j$ because $w_{j,p}$ facilitates calculation of the flux density distribution ${\bf j}(\mathbf{x},\mathbf{v},t)$.

Simulating macroparticles following Eqs.~(\ref{eq:slpicEOMx}--\ref{eq:slpicEOMw}), we can compute the phase-space density distribution function $f$ at any time via Eq.~(\ref{eq:Ansatz}). However, with SLPIC it is often especially useful to consider particle fluxes. We can compute the flux density distribution, $\mathbf{j}(\mathbf{x}, \mathbf{v}, t)=\mathbf{v} f(\mathbf{x}, \mathbf{v}, t)$, as follows:
\begin{eqnarray} \label{eq:j}
  \mathbf{j}(\mathbf{x}, \mathbf{v}, t) 
  =
  \sum_p \mathbf{v}_p w_p(t) S[\mathbf{x}-\mathbf{x}_p(t)] 
             \delta^3 [\mathbf{v}-\mathbf{v}_p(t)]
  =
  \sum_p \beta(v_p) \mathbf{v}_p w_{j,p} S[\mathbf{x}-\mathbf{x}_p(t)] 
             \delta^3 [\mathbf{v}-\mathbf{v}_p(t)].
  \nonumber \\
\end{eqnarray}
To estimate the flux $\mathbf{j}\cdot\hat{\bf n}dA$ through some surface element $\hat{\bf n}dA$ from the macroparticles crossing it in time $\Delta t$, we consider that a macroparticle with $\dot{\bf x}$ will cross the surface if it is within a distance $\dot{\bf x} \Delta t \cdot \hat{\bf n}$ of the surface. From Eq.~(\ref{eq:j}) we see that, since $\dot{\bf x}=\beta(v_p)\mathbf{v}_p$, the flux (integrated over time~$\Delta t$) through the surface element is simply the sum of the $w_{j,p}$ over all macroparticles crossing the surface in $\Delta t$.

Because the $w_{j,p}$ are constant in time, a macroparticle always represents the same physical flux whenever it crosses a surface.  If a macroparticle~$p$ has $w_{j,p}=100$, then every time it crosses a surface, it represents 100 physical particles crossing the surface. This statement is trivial in PIC, but in SLPIC, where $w_p(t)$ varies in time, it is nontrivial and very useful. For example, it ensures that, if a macroparticle enters into some volume $V$ and later exits that volume, the time-integrated flux through the surface (due to that particle) is zero.

At this point readers might suspect a paradox in the SLPIC treatment arising from a difference between~$w_p$ and~$w_{j,p}$.  Consider a macroparticle with $v_p=10v_0$, $\beta(v_p)=0.1$, $w_{j,p}=100$, and $w_p=10$, and suppose it crosses a surface to enter volume~$V$ during time interval~$\Delta t$. That macroparticle represents $w_p=10$ physical particles in volume~$V$; however, it represented $w_{j,p}=100$ physical particles crossing the surface into~$V$. This counterintuitive behavior is correct for SLPIC. It is counterintuitive because PIC has led us to assume that a macroparticle with $w_p$ and $(\mathbf{x}_p, \mathbf{v}_p)$ represents a swarm of $w_p$ physical particles within the volume $d^3\!x\,d^3\!v$ around $(\mathbf{x}_p, \mathbf{v}_p)$, and that those physical particles travel roughly with the macroparticle to $(\mathbf{x}_p+\dot{\bf x}_p\Delta t,\mathbf{v}_p+\dot{\bf v}_p\Delta t)$ over time $\Delta t$. Fundamentally, however, a macroparticle represents a chunk of the distribution $f d^3\!x\, d^3\!v$, and SLPIC decouples macroparticles from the physical particles they represent. A speed-limited SLPIC macroparticle moves in phase space with $\dot{\bf x}_p$ and $\dot{\bf v}_p$ slower than the physical particles it represents. Thus the same macroparticle may represent one set of physical particles at time~$t$ and a different set of particles at time~$t+\Delta t$.

In the above example, the density represented by the SLPIC macroparticle corresponds to $w_p=10$ physical particles that are near $[\mathbf{x}_p(t),\mathbf{v}_p(t)]$ at time $t$; however, over time interval~$\Delta t$, the physical flux represented by the macroparticle includes all the physical particles that would would be near $[\mathbf{x}_p(t'),\mathbf{v}_p(t')]$ at any time~$t'\in[t,t+\Delta t]$. Since the physical particles travel 10 times faster than the macroparticle, the physical flux is $10w_p(t)/\Delta t = w_{j,p}/\Delta t$.

It is important to remember that SLPIC is accurate only if $f(\mathbf{x},\mathbf{v},t)$ changes sufficiently slowly---and it is as accurate as PIC in the steady-state limit. The $f(\mathbf{x},\mathbf{v},t)$ describing a single particle with $v>v_0$ is not a slowly-changing function; therefore, one typically cannot verify SLPIC based on single-particle thought experiments. In the above single-particle example, the continuity equation is violated because~$f(\mathbf{x},\mathbf{v},t)$ changes too rapidly for the SLPIC approximation to be valid.

Exactly what constitutes ``sufficiently slowly'' for SLPIC accuracy remains an open question.  We have previously observed that for (Langmuir) wave-particle interaction in~1D, ``sufficiently slowly'' means that the wave phase velocity must be slower than the SLPIC speed limit for accurate simulation \citep{werner2018speeding,jenkins2021Dispersion}. Similarly, although studying SLPIC in magnetized plasmas is beyond the scope of this work, we imagine that the electron-cyclotron resonance interaction will be too fast for accurate simulation, as long as SLPIC reduces the electron gyrofrequency---but ion-cyclotron resonance might be accurately simulated.  Although disadvantageous, it is precisely because SLPIC neglects fast motions that it can offer faster simulation.
Importantly, in this paper we apply SLPIC to a steady-state problem in the sense that in the very beginning of breakdown, the space charge has negligible effect on the applied constant field; in the steady-state limit, we expect SLPIC to be essentially as accurate as PIC.

Reiterating an important point: the flux weight $w_{j,p}=w_p(t)/\beta(v_p(t))$ of a SLPIC macroparticle is a constant in time, and this ensures that the flux density is divergenceless in the steady-state limit. For example, if a particle receives an abrupt kick---whether from the electric field or from a collision---its $w_{j,p}$ remains constant.  If the impulse increases the particle's speed $v_p$, its density weight $w_p$ must change accordingly so that $w_p = \beta(v_p) w_{j,p}$. In Sec. \ref{sec:collisions}, we will see that the flux weight is conserved when simulating collisions involving SLPIC particles.

\section{Collisions in SLPIC} \label{sec:collisions}
Collisions can be simulated in PIC with a PIC-MCC algorithm;\cite{birdsall1991particle,bird1994molecular,nanbu2000probability} the MCC algorithm can be used for SLPIC with just a few modifications: (1) using the flux weight $w_{j,p}$ instead of the density weight $w_p$ (cf.~\S\ref{sec:slpicIntro}), and (2) considering the slowing of time when determining the collision rate. (Since most SLPIC simulations will have macroparticles with a distribution of~$w_{j,p}$, the PIC-MCC algorithm must be able to handle macroparticles with different weights before modification for SLPIC.)

Before discussing collisions in SLPIC, we briefly review the pertinent aspects of collisions in PIC. In PIC-MCC, a binary macroparticle collision is not at all the same as a binary collision between two physical particles. The collision between two macroparticles must statistically represent collisions between the two swarms of physical particles represented by the macroparticles. The spatial distribution of physical particles is usually assumed to be uniform within the same grid cell containing the macroparticle centers, regardless of the macroparticle shape $S$ [cf.~Eq.~(\ref{eq:Ansatz})], and zero outside that cell.

For concreteness, let us consider electron impact ionization in PIC-MCC, in which a primary electron collides with a primary neutral atom, resulting in the scattering of the primary electron, the neutral ``becoming'' an ion, and the ``creation'' of a secondary electron.  (Although there are other kinds of collisions, e.g., elastic or excitation, between electrons and neutrals that do not ionize, we consider only ionizing collisions for simplicity.)
A primary electron macroparticle $p$ with weight $w_{p0}$ and velocity $\mathbf{v}_{p0}$ collides with a neutral macro-atom with $w_{n0}$ and $\mathbf{v}_{n0}$. The primary electron macroparticle will be split into two parts: one part with weight $w_{p1}$ represents uncollided electrons and the other with weight $w_{p1}'$ represents scattered primary electrons (recoiling from having just ionized atoms).  Similarly, $w_{n1}$ is the weight of uncollided neutrals, and $w_{i}$ the weight of ``scattered'' neutrals, which are really ions and are now subject to acceleration in the applied electric field. In addition, secondary electrons are created with weight~$w_{se}$.
In principle there could be many scattered/secondary macroparticles with different velocities, but for simplicity we will assume at most one per primary species.
The uncollided particles retain the original velocities; the other particles must be assigned new velocities according to the ionization process: $\mathbf{v}_{p1}'$, $\mathbf{v}_{i}$, and $\mathbf{v}_{se}$.
Conservation of subatomic particles requires $w_{p0}=w_{p1}+w_{p1}'$, $w_{n0}=w_{n1}+w_{i}$, and $w_{p1}'=w_{i}=w_{se}=W_c$, where $W_c$ is the number of (physical-particle) collisions expected to take place when the two macroparticles ``collide.''

The expected number of collisions $W_c$ within a time interval $\Delta t$ in a cell volume $\Delta V$  between two arbitrary particle distributions, $f_p$ (primary electrons) and $f_n$ (neutrals) is
\begin{eqnarray} \label{eq:collisionNumber}
  W_c &=& \Delta V \Delta t \int d^3 
    \mathbf{v}_{p0} \int d^3 \mathbf{v}_{n0}  \,
    f_{p}(\mathbf{v}_{p0})
    f_{n}(\mathbf{v}_{p0})
    \sigma(\mathbf{v}_{n0}-\mathbf{v}_{p0})
    \| \mathbf{v}_{n0}-\mathbf{v}_{p0} \|
\end{eqnarray}
where $\sigma(\mathbf{v}_r)$ is the collision cross-section, which depends on the relative velocity $\mathbf{v}_r$ between colliding particles. To calculate $W_c$ between two PIC macroparticles in the same grid cell, we substitute the appropriate distributions, i.e., $f_p(\mathbf v) = (w_{p0}/\Delta V) \delta^3(\mathbf{v} - \mathbf{v}_{p0})$ and $f_n(\mathbf{v})=(w_{n0}/\Delta V)\delta^3(\mathbf{v} - \mathbf{v}_{n0})$, ignoring the shape functions of the macroparticles, treating them as uniform in one cell and zero outside. The expected number of collisions between the two macroparticles is then
\begin{eqnarray} \label{eq:PICcollisionNumber}
  W_{c} &=& \Delta V 
    \frac{w_{p0}}{\Delta V}
    \frac{w_{n0}}{\Delta V}
    \sigma(\mathbf{v}_{n0}-\mathbf{v}_{p0})
    \| \mathbf{v}_{n0}-\mathbf{v}_{p0} \|\Delta t 
.\end{eqnarray}
This can be verified by considering a single electron moving through a swarm of atoms of density $n_{n0}=w_{n0}/\Delta V$ with velocity $\mathbf{v}_{n0}$. In the reference frame co-moving with the atoms, $d= \| \mathbf{v}_{n0}-\mathbf{v}_{p0} \|\Delta t $ is the distance traveled by the electron in $\Delta t$, and $\lambda_{\rm mfp}= (n_{n0} \sigma)^{-1}$ is its mean free path. Therefore $d/\lambda_{\rm mfp}$ is the expected number of collisions that a single electron would experience (for small $\Delta t$ such that $d/\lambda_{\rm mfp} \ll 1$). The total number of electron collisions is then $W_c=w_{p0} d/\lambda_{\rm mfp}$, which is identical to Eq.~(\ref{eq:PICcollisionNumber}).

There are multiple Monte Carlo strategies for colliding pairs of macroparticles that all yield the same $W_c$ \emph{on average}. For example, if $W_c = 0.01 w_{p0}$, we could avoid splitting the primary macroparticles and creating small-weight scattered/secondary particles by ignoring~99\% of colliding pairs, and colliding~1\% with an enhanced $W_c'=w_{p0}$. We will not discuss this in any detail, because these strategies are identical for PIC and SLPIC. We also omit other practical details that need no modification for SLPIC, such as determining scattered and secondary macroparticle velocities according to given differential cross-sections, and dealing with exceptional cases such as $W_c > w_{p0}$ (indicating that $\Delta t$ is too large). The only differences introduced by SLPIC are the determination of $W_c$ and the resulting macroparticle weights.

As described in~\S\ref{sec:slpicIntro}, in SLPIC it is better to verify physical accuracy by considering particle fluxes rather than numbers or densities. Thus the basic thought-experiment for developing SLPIC collision algorithms is not the collision of two macroparticles, but of two narrow beams of macroparticles intersecting in a small volume; the particle fluxes into and out of that volume must satisfy the appropriate conservation laws. For electron impact ionization, this means that $w_{j,p0}=w_{j,p1}+w_{j,p1}'$, $w_{j,n0}=w_{j,n1}+w_{j,i}$, and $w_{j,p1}'=w_{j,i}=w_{j,se}=W_{j,c}$. Because flux weights $w_j$ are constant in time [unlike density weights $w_p(t)$], this ensures that the fluxes of particles entering and exiting $\Delta V$ are appropriately perserved; i.e., the flux of unscattered plus scattered primaries leaving $\Delta V$ equals the flux of primaries entering $\Delta V$, and the fluxes of secondary electrons and scattered primaries leaving $\Delta V$ are equal, etc.

The rule for determining the collision rate, $W_{j,c}/\Delta t$, is that SLPIC preserves the physical mean free path. For example, consider a steady-state, monoenergetic beam of particles in the~$+x$ direction, entering some scattering medium with mean free path~$\lambda_{\rm mfp}$. Scattering causes the density of original beam particles to decrease as~$\sim \exp(-x/\lambda_{\rm mfp})$. Because this is a steady-state, SLPIC should render this density profile accurately, requiring that SLPIC macroparticles experience the physical mean free path. In other words, SLPIC macroparticle must scatter with probability $d'/\lambda_{\rm mfp}=\dot{x}\Delta t/\lambda_{\rm mfp} = \beta(v) v\Delta t/\lambda_{\rm mfp}$ within time $\Delta t$ (in contrast, a physical particle would scatter with probability $d/\lambda_{\rm mfp}=\dot{x}\Delta t/\lambda_{\rm mfp} = v\Delta t/\lambda_{\rm mfp}$). This can also be viewed as the reduction in collision rate due to the slowing down of time by a factor $\beta(v)$. The collision rate (in terms of fluxes) in $\Delta V$ must therefore be
\begin{eqnarray}
  \frac{W_{j,c}}{\Delta t} &=& 
    \frac{w_{j,p0}}{\Delta t}
    \frac{d'}{\lambda_{\rm mfp}}
    =
    \Delta V  
    \frac{w_{j,p0}/\Delta t}{\Delta V}
    n_{n0} \sigma(\mathbf{v}_{n0}-\mathbf{v}_{p0})
    \| \mathbf{v}_{ni}-\mathbf{v}_{pi} \| \beta_p(v_{p0}) \Delta t
\end{eqnarray}
where $\beta_p(v)$ is the speed-limiting function for electrons. This merely says that---if one were to measure fluxes into and out of $\Delta V$---the flux of scattered electrons (i.e., $W_{j,c}/\Delta t$) would equal the flux of incident electrons (i.e., $w_{j,p0}/\Delta t$) times $d'/\lambda_{\rm mfp}$. 

The above expression involves the neutral density: $n_{n0} = w_{n0}/\Delta V = \beta_n (v_{n0}) w_{j,n0} / \Delta V $. (In most applications, atoms will be slow and not speed-limited, hence $\beta_n\equiv 1$, but we want this treatment to extend to arbitrary collisions.) Thus:
\begin{eqnarray} \label{eq:SLPICcollisionNumber}
  W_{j,c} &=& 
    \Delta V  
    \frac{w_{j,p0}}{\Delta V}
    \frac{\beta_n(v_{n0}) w_{j,n0}}{\Delta V}
    \sigma(\mathbf{v}_{n0}-\mathbf{v}_{p0})
    \| \mathbf{v}_{n0}-\mathbf{v}_{p0} \| \beta_p(v_{p0}) \Delta t
    \nonumber \\
    &=&
    \Delta V  
    \frac{w_{j,p0}}{\Delta V}
    \frac{w_{j,n0}}{\Delta V}
    \sigma(\mathbf{v}_{n0}-\mathbf{v}_{p0})
    \| \mathbf{v}_{n0}-\mathbf{v}_{p0} \| 
      \beta_n(v_{n0}) \beta_p(v_{p0}) \Delta t
.\end{eqnarray}
This expression (or rather $W_{j,c}/\Delta t$) yields the number of physical electrons scattered per $\Delta t$ from the collision of particles represented by two SLPIC macroparticles. It is the same as Eq.~(\ref{eq:PICcollisionNumber}), except: (1) flux weights $w_j$ are used (not density weights $w$), and (2) the timestep is modified by a factor $\beta_n(v_{n0}) \beta_p(v_{p0})$.

After calculating $W_{j,c}$ using Eq.~(\ref{eq:SLPICcollisionNumber}), the rest of the collision algorithm proceeds as in PIC, except that where a weight was used in the PIC algorithm, the flux weight must be used in SLPIC.  Once a macroparticle's flux weight $w_{j,p}$ is determined, its density weight is set by $w_p = \beta(v)w_{j,p}$.

\section{Simulation Setup and Methodology} \label{sec:method}
We modelled an argon-filled parallel plate capacitor with one spatial dimension and three velocity dimensions. The electrodes were treated as particle-absorbing boundaries and the gap-distance was fixed at $d=1$~cm. We imposed a constant electric field $E=V/d$. The cell size $\Delta x$ was set to a quarter of the mean free path. The number of cells for each simulation is given in Table~\ref{tab:sims}. We ran both SLPIC and PIC simulations. The setup parameters for the SLPIC and PIC simulations were identical except for the timestep. For PIC, we used a timestep given by $\Delta t = \Delta x/v_{\rm max}$, where $v_{\rm max}=\sqrt{2eV/m_e}$ was the maximum electron speed. For SLPIC, we used a timestep given by $\Delta t = \Delta x/v_{0}$, where $v_{0}=v_{\rm max}\sqrt{m_e/m_{Ar}}$ was the electron speed limit imposed by $\beta(v_p)$ given in Eq.~(\ref{eq:betaAbrupt}). This $v_0$ was chosen because it provided the maximum speed up without affecting the ion motion. The large mass difference between electrons and argon ions resulted in the SLPIC timestep being 270 times larger than the PIC timestep. Therefore, the SLPIC simulations required 270 times fewer timesteps than PIC. The number of timesteps for each simulation is given in Table~\ref{tab:sims}.

To initiate the discharge we injected 100 electrons at the cathode in the first timestep with zero initial velocity. We simulated five types of electron-neutral collisions: ionization, elastic collisions, and three excitations. Many different types of collisions occur between electrons and argon, but we simulated only the five most likely. The other collisions had negligible cross-sections in the relevant electron energy range. The cross-sections were extracted from the Biagi-v7.1 database\cite{biagi1999monte} and Phelps database\cite{yamabe1983measurement} on www.lxcat.net on August 15, 2020. We did not include any ion collisions. Although ion collisions could increase $V_b$, $\gamma_{se}$ depends only weakly on the ion energy, so the effect of these collisions is negligible. For excitation collisions, the electrons were scattered isotropically with an energy reduced by the excitation threshold. For elastic collisions, the electrons were scattered according to the Vahedi-Surendra algorithm.\cite{vahedi1995monte} For ionization collisions, the products were generated according to an algorithm developed by \citet{kutasi2000hybrid}. We used an energy-dependent model of secondary electron emission due to ion impact at the cathode.\cite{phelps1999cold} Secondary electrons were emitted from the cathode with zero velocity with a probability given by Eq.~(\ref{eq:sey}) that depends weakly on the energy of the incident ion $\epsilon_i$.
\begin{equation} \label{eq:sey}
\gamma_{se}=
\begin{cases} 
      0.09\left(\frac{\epsilon_i}{700\,\rm{eV}}\right)^{0.05} ,& \epsilon_i < 700~\rm{eV} \\
      0.09\left(\frac{\epsilon_i}{700\,\rm{eV}}\right)^{0.72} ,& \epsilon_i \geq 700~\rm{eV} \\
\end{cases}
\end{equation}

To determine whether the simulation voltage $V$ was above or below $V_b$, we tracked the ion population in the simulation over 30 ion crossing times ($30\sqrt{2m_{Ar}d/eE}$). For $V<V_b$, the ion population decreases with time after an initial rise, eventually returning to zero. For $V>V_b$, the ion population increases with time. This criterion is equivalent to the breakdown condition given by $\gamma_{se}(e^{\alpha d}-1) > 1$. This method provided a clear binary classification method of simulation results. For each pressure, we performed a grid search in the vicinity of the experimental $V_b$. Once a simulation above and a simulation below breakdown were found, we bracketed the interval with the respective voltages. These brackets are indicated by the bars in Fig.~\ref{fig:paschen}. 

\begin{figure}
\includegraphics{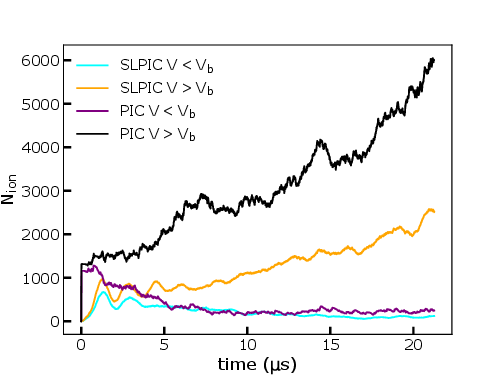}
\caption{\label{fig:method}The number of ions in the simulation domain as a function of time for the simulations with $V$ nearest $V_b$ at $pd=1$~Torr~cm. The SLPIC results are given for $V>V_b$ (orange) and $V<V_b$ (cyan). The PIC results are given for $V>V_b$ (black) and $V<V_b$ (purple). The number of ions in the simulation domain increases over time when $V>V_b$, and decreases when $V<V_b$.}
\end{figure}

The evolutions of the ion population for simulations above and below breakdown at 1~Torr~cm are shown in Fig.~\ref{fig:method}, for both SLPIC and PIC. It can be seen that for $V>V_b$, the ion population increases with time, and for $V<V_b$, the ion population decreases with time after the initial rise. While SLPIC and PIC give the same results for $V_b$, the dynamics of the simulation are clearly different. The initial rise in ion population is steeper for PIC because the seed electrons cross the simulation domain and ionize the argon gas almost immediately, while in SLPIC, the speed-limited seed electrons cross the domain more slowly and therefore take longer to ionize the argon gas. The speed limiting also causes the ion population to evolve more smoothly as electrons and ions now have comparable velocities and oscillate in and out of the simulation domain at similar frequencies, but out of phase. SLPIC cannot capture the transient behavior of the particles since the speed-limiting approximation Eq.~(\ref{eq:slpicApprox}) is not satisfied in this regime. However, we do see convergence of the SLPIC dynamics to PIC as we increase the speed-limit.

We bracketed the breakdown voltage for 11 values of $pd$ ranging from 0.3 to 300~Torr~cm. We performed simulations with the Vorpal\cite{nieter2004vorpal} code distributed in VSim-11. The simulations ran on the CORI supercomputer at the National Energy Research Scientific Computing Center (NERSC). The number of cores for each simulation is given in Table~\ref{tab:sims}. 

\section{Simulation Results} \label{sec:results}
\begin{table*}[ht]
\caption{\label{tab:sims}Summary of setup parameters and performance for PIC and SLPIC simulations with $V \approx V_b$. The gap-distance was fixed at $d=$1~cm. The number of timesteps is given by $N_T$.``Cores" refers to the number of CORI cores that the simulation was run on. The runtime $T$ is given in core-hours. The speed-up is given by $T_{\rm PIC}/T_{\rm SLPIC}$.}
\begin{ruledtabular}
\begin{tabular}{ccccccccc}
$p$ (Torr) & $V$ (V) & Cells & $N_{T,\rm SLPIC}$ & $N_{T,\rm PIC}$ & Cores & $T_{\rm SLPIC}$ & $T_{\rm PIC}$ & Speed-up \\
\hline 
0.3 & 260 & 66 & 4.0$\times 10^{3}$ & 1.1$\times 10^{6}$ & 4 & 0.26 & 11 & 42 \\
0.4 & 195 & 88 & 5.3$\times 10^{3}$ & 1.4$\times 10^{6}$ & 4 & 0.27 & 14 & 52 \\
0.6 & 175 & 131 & 7.9$\times 10^{3}$ & 2.1$\times 10^{6}$ & 4 & 0.49 & 91 & 187 \\
1 & 170 & 219 & 1.3$\times 10^{4}$ & 3.5$\times 10^{6}$ & 4 & 0.78 & 113 & 146 \\
2 & 185 & 437 & 2.6$\times 10^{4}$ & 7.1$\times 10^{6}$ & 4 & 1.8 & 91 & 51 \\
4 & 225 & 874 & 5.2$\times 10^{4}$ & 1.4$\times 10^{7}$ & 4 & 6.3 & 1308~\footnote{time based on extrapolation due to time restrictions on CORI} & 208 \\
6 & 262 & 1310 & 7.9$\times 10^{4}$ & N/A & 4 & 9.2 & N/A & N/A \\
10 & 340 & 2184 & 1.3$\times 10^{5}$ & N/A & 4 & 13 & N/A & N/A \\
30 & 620 & 6550 & 3.9$\times 10^{5}$ & N/A & 128 & 105 & N/A & N/A \\
100 & 1400 & 21832 & 1.3$\times 10^{6}$ & N/A & 128 & 771 & N/A & N/A \\
300 & 3100 & 65494 & 3.9$\times 10^{6}$ & N/A & 128 & 6400~$^{\rm{a}}$ & N/A & N/A \\
\end{tabular}
\end{ruledtabular}
\end{table*}

\begin{figure}
\includegraphics{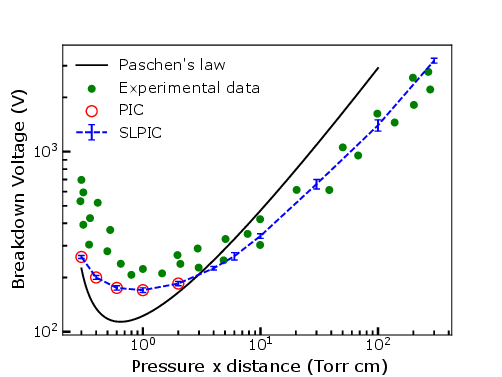}
\caption{\label{fig:paschen}The Paschen curve for argon. The experimental data (solid, green circles) was extracted from \citet{phelps1999cold}. The values of $A$, $B$, and $\gamma_{se}$ in Paschen's law (solid, black line) were taken from \citet{lieberman2005principles}. The SLPIC results (blue bars connected by dashed, blue line) display the upper and lower brackets on the simulated breakdown voltage. The 5 lowest pressures (open, red circles) were also simulated using PIC, which yielded the same bounds as SLPIC.}
\end{figure}

SLPIC accurately computed the Paschen curve for argon over three orders of magnitude in $pd$, 0.3 to 300~Torr~cm, agreeing with PIC over the range where PIC was feasible, 0.3 to 2~Torr~cm. For each value of $pd$, we bracketed the breakdown voltage with one simulation above and below breakdown. The Paschen curve generated by the SLPIC and PIC simulations is shown along with Paschen's law and experimental data from \citet{phelps1999cold} in Fig.~\ref{fig:paschen}. For the values of $pd$ where PIC was run, PIC and SLPIC classified each simulation identically, thus yielding the same brackets. Below 3~Torr~cm, the simulated breakdown voltages were lower than those measured in experiment. At low $pd$ the experiments approach the vacuum discharge regime where processes such as out-gassing, vacuum arc by burned cathode, and flashover become relevant to breakdown.\cite{descoeudres2009investigation,sun2018simulation} We did not simulate these effects, which may account for the discrepancy. Paschen's law, given in Eq.~(\ref{eq:paschen}), is plotted using the values of $A=11.5$~cm$^{-1}$~Torr$^{-1}$, $B=176$~V~cm$^{-1}$~Torr$^{-1}$, and $\gamma_{se}=0.07$ taken from \citet{lieberman2005principles}. Below 0.3~Torr~cm, the simulation did not break down. This matches Paschen's law [Eq.~(\ref{eq:paschen})] which exhibits a singularity below 0.3~Torr~cm.

SLPIC ran 40 to 200 times faster than PIC. The simulation parameters and performance for a subset of simulations run near the breakdown voltage are given in Table~\ref{tab:sims}. The speed-ups are the ratios of the PIC runtimes to the SLPIC runtimes and are given in the rightmost column. On average, the SLPIC simulations in Table~\ref{tab:sims} ran 116 times faster than PIC. Above 4~Torr~cm, it became unfeasible to run PIC simulations. SLPIC enabled us to explore these higher pressures with a fully kinetic treatment.

\section{Summary}
Fast, accurate electrical discharge simulations are needed for the design of plasma processing equipment, sensitive microchips, and volatile chemical processing facilities. We have demonstrated that SLPIC is as accurate as PIC, but faster in predicting breakdown voltages of a gas-filled capacitor. SLPIC accurately computed the Paschen curve $V_b(pd)$ for argon for $pd$ ranging from 0.3 to 300~Torr~cm. SLPIC and PIC produced identical results, but SLPIC ran 40 to 200 times faster and extended the range of feasible simulations. In accurately computing the Paschen curve for argon, SLPIC has demonstrated that it can accurately model collisions, including electron-neutral ionization, electron-neutral elastic collisions,  electron-neutral excitations, and ion-induced secondary electron emission. We expect that SLPIC will also be as accurate as PIC, but much faster, in simulating voltage breakdown in gases in more complicated 3D geometries as well as in the glow discharge regime where the discharge current alters the applied electric field.

\begin{acknowledgments}
This work was supported by NSF grant PHY1707430 and by U.S. Department of Energy SBIR Phase II Award DE-SC0015762. This research used resources of the National Energy Research Scientific Computing Center (NERSC), a U.S. Department of Energy Office of Science User Facility operated under Contract No. DE-AC02-05CH11231.
\end{acknowledgments}
\section*{AIP Publishing Data Sharing Policy}
The data that support the findings of this study are available from the corresponding author upon reasonable request.
\bibliography{slpicPaschen}
\end{document}